\documentclass[conference]{IEEEtran}
\usepackage{cite}
\usepackage{amsmath,amssymb,amsfonts}
\usepackage{algorithmic}
\usepackage{graphicx}
\usepackage{textcomp}
\usepackage{xcolor}
\usepackage{algorithm}
\usepackage{nicefrac}
\usepackage{footmisc}

\usepackage{footnote}
\makesavenoteenv{tabular}
\makesavenoteenv{table}

\newcommand{\comment}[1]{}
\def\BibTeX{{\rm B\kern-.05em{\sc i\kern-.025em b}\kern-.08em
    T\kern-.1667em\lower.7ex\hbox{E}\kern-.125emX}}
\begin{document}

\title{Quantum-resistant digital signatures schemes for low-power IoT}

\author{\IEEEauthorblockN{1\textsuperscript{st} Hannes Hattenbach}
\IEEEauthorblockA{\textit{Computational Science} \\
\textit{Freie Universität}\\
Berlin, DE \\
hannes.hattenbach@fu-berlin.de}
}

\maketitle

\begin{abstract}
Quantum computers are on the horizon to get to a sufficient size that will then be able to break pretty much all the encryption and signature schemes we currently use. This is the case for human interface devices as well as for IoT nodes. 
In this paper i am comparing some signature schemes currently in the process of standardization by the NIST.
After explaining the underlying basis on why some schemes are different in some aspects compared to others i will evaluate which currently available implementations are better suited for usage in IoT use-cases.
We will come to further focus on the most promising schemes FALCON and Dilithium, which differ in one signifiant aspect that makes FALCON worse for signing but very good for verification purposes.
\end{abstract}



\section{Introduction}
\comment{ 
} 

The quantum revolution is coming. With quantum computers\footnote{compare section \ref{l:quantum_computing}} on the way to get more and more functional, people are fearing a loss of their security and privacy.
Or as \cite{QR_sigs} puts it, ``principles of data integrity, message authentication, and nonrepudiation, are going to have profound aftermath on sensory data in terms of security and privacy.''
That is because there are algorithms based on Shors algorithm that can forge signatures and decrypt encrypted messages whose security is based on discrete logarithms, including elliptic curves or prime factorization, like our most common schemes Elliptic Curve Digital Signature Algorithm (ECDSA) and RSA respectively are.
The quantum computer only needs access to the public keys of these asymmetric schemes two forge the private keys and therefor decipher all encrypted messages or forge arbitrary signatures.
The expenditure to forge a signature\footnote{that is considered secure under normal circumstances 
} with classic\footnote{we refer to classic if something is not directly leveraging entanglement or superposition} computers rises exponentially with increased key length, therefor being essentially unbreakable by classic computers.
A sufficient quantum computer on the other hand can derive a private key from a public key in polynomial time, therefor rendering these schemes broken.

That is why there are currently schemes under standardization\cite{PQClean-GH,nist_finalists_website} that are based on other hard problems (not number theory) like so called lattice problems that cannot be that easily forged by quantum computers to save our privacy and security.

One of the use cases not directly coming to mind for the end user, but being as important non the less is signing sensitive sensor data in the Internet of Things (IoT).
Another problem coming up in the IoT compared to end-user-devices like Laptops and Smartphones though is the severe resource constraint-ness. 
The IoT consist of low power devices with very few storage and computing power.

In this paper i am going to evaluate existing signature schemes and their usage possibilities for the IoT regarding their performance metrics.

Therefor i am going to give a small introduction and background to quantum computing, being a little more detailed about their ability to break current encryption and signature standards as well as on the internet of things.
In the next section i will give an overview over current candidates for Quantum Resistant (QR) Algorithms and giving performance metrics for those.
I will give an overview on what kind of underlying mathematical problems QR algorithms rely on, with focus on the two most used kinds: lattice and hash based schemes. 
The following chapter will then focus on signature schemes in the IoT, starting with additional performance metrics relevant in the IoT.
And finally i will be focussing on the best signature contenders for the IoT so far: FALCON and Dilithium.

\section{Background}\label{background}
\subsection{Cryptography}\label{bg:crypto}
Loosely speaking the main topic of cryptography can be divided into three groups.
The first of these groups is about one way functions, that shall not, as the name implies, be efficiently reversible.
If we create a smaller value of constant length from a bigger set of possibly variable length, we commonly refer to that as \textit{hashing}.
Cryptographic hashing is important for a variety of different applications like storing and matching passwords without the ability to infer any knowledge about that password.
Hashing itself can be used for the next pillar of cryptography: signatures.
Signature schemes are used to proof integrity or authenticity of any data.
A signature scheme consists of two parts, \textit{signing} and \textit{verifying}. 
The last group is encryption, which ensures privacy/confidentiality of any data, s.t. only the right entities can decrypt this data.
These schemes consist of the two parts \textit{encryption} and \textit{decryption}.
Additionally to those parts for signatures as well as encryption there needs to be process of \textit{key-generation}.
We also differentiate between symmetric and asymmetric schemes. The first one has a different private and public key while the latter uses the same for de- and encryption.
More details about which of those schemes will be more or less endangered by quantum computing are in section \ref{l:quantum_computing} and \ref{l:qr-algs}.
\newline
In general we denote a signature scheme as the group of three algorithms \{GEN, SIGN, VER\} and a encryption scheme as \{GEN, ENC, DEC\}.

\subsection{Internet of Things}\label{bg:iot}
\comment{ 
} 

The IoT consists of a growing number (currently over 3 billion \cite{QR_IoT}) of devices of all sorts, having in common, that they communicate with each other and the environment rather than directly with humans.
Those devices range from automatic lights and smart home devices to tiny interconnected sensors in automatic fabrication.
A common characteristic though is, that most of these devices have limited processing power, flash storage and random access memory (RAM). 
A popular example for hobbyist IoT devices is the ESP32 from Espressif Microsystems.
They offer multiple modules with up to 240Mhz clock on the 32bit IC, up to 16MiB Flash Storage and 320KiB RAM.
Which is more than other comparable devices but way less then a lower spec modern smartphone, with 10 times the frequency, 4GB of RAM and 64GB of storage.

Since the IoT consists of very different types of constrained nodes the IETF introduced different classes on which to classify IoT nodes, those can be seen in table \ref{IoT-classes}

\begin{table}
    \label{IoT-classes}
    \centering
    \caption{IETF IoT Classes}
    \begin{tabular}{|l | c c|}
        \hline
        Class & RAM & Flash \\
        \hline
        C0 & $<<$ 10 KiB & $<<$ 100 KiB\\
        C1 & ~ 10 KiB & ~ 100 KiB\\
        C2 & ~ 50 KiB & ~ 250 KiB\\
        \hline
    \end{tabular} 
\end{table}

\section{Quantum Resistant Security}
\subsection{Quantum Computing}\label{l:quantum_computing}
In contrast to classical computers, where information is processed in discrete states, a quantum computer leverages quantum mechanics to operate on so-called qubits - quantum objects that can be in superposition or entangled with each other. 
Opening a new kind of computing. 
One of the implications of that is, that it is now possible to factor large numbers in polynomial time using an algorithm developed by Shor \cite{Shor}. 
This algorithm uses a so-called Quantum-Fourier-Transform (QFT) to (probabilistically) get the frequencies of which a given function output occurs. That can be used together with euclids algorithm of finding the greatest common devisor to derive the prime factors. 
Prior to to quantum computers this was considered a hard problem that could only be computed in exponential time and was therefor considered practically impossible and was used as the basis-problem for RSA encryption.
Similar to that other common schemes like ECDSA can also be broken be slightly modified versions of Shors Algorithm.
\subsection{QR Algorithms}\label{l:qr-algs}
The two main algorithms with practical use cases that have a great speed-up compared to classical solutions, are the already introduced algorithm by Shor and an algorithm by Grover that can essentially reverses one-way functions by creating a superposition over all possible inputs, flipping all inputs with the wanted output (without knowing the inputs) and then flipping this state about its mean and repeating this process a lot of times \cite{Grover}.
While Shors algorithm provides exponential speed-up, Grovers algorithm only provides quadratic speed-up. It was also shown, that something similar to grovers algorithm but with exponential speedup is impossible \cite{Strengths&Weaknesses_QC}. 
Which implies that hashing as well as symmetric cryptography stays relatively secure.
The quadratic speedup provided by quantum computers can easily be mitigated by doubling the key length.
On the other hand though, classical asymmetric cryptography is endangered by shors algorithm and quantum computers.

But not all asymmetric cryptography schemes are equally affected.
There are different proposals, both for QR encryption and for QR signature schemes.
They all do have in common though, that their security is not absolutely mathematically proven, but based upon assumptions.
We therefor need to consider a few measures that make schemes more or less secure.

\subsection{Performance Metrics}
\comment{ 
\cite{springer_security_analyses}:
Qubit cost, Quantum Gate cost
} 

Some performance metrics exist in QR schemes as well as in classic schemes.

Key length and key exchange message length \cite{QR_algs} are the more obvious ones.
The computing time also comes to mind as a performance metric. Here you need to differentiate between key generation, which is less important, since it should only occur rarely, and signing as well as signature verification \footnote{as well as its counterparts de- and encryption}.

Primarily in signatures another metric arises: how often can a private key be used before it needs to be switched out for another one, because the signature leaked information of the key.
This is not particularly relevant in most cases, as methods can be used to create long term procedures from short term procedures (those where a key can rarely, if ever, be recycled).
But it is relevant in the case of the IoT, since those methods require extra memory which is sparse in IoT-devices. Additionally they tend to make the signatures themselves longer, which also is not preferable in the IoT. \cite{QR_algs}

Additionally to more traditional performance metrics we somehow need to measure the security of given schemes against an attack by a quantum computer.

The main measurement we can take to measure quantum resistance is to count how many quantum gates or quantum bits are needed.\cite{springer_security_analyses}

But unfortunately this is rather hard and there is currently no standard benchmark to measure quantum resistance \cite{QR_comparison}, nevertheless the National Institute of Science and Technology (NIST) created a standard that describes how secure a scheme is against a quantum computer by classifying it within 5 classes that can be determined with grovers algorithm \cite{QR_Iot_Lattice,Energy_comp}.
Those classes can be seen in table \ref{QR-classes}.

\begin{table}
    \label{QR-classes}
    \centering
    \caption{QR Security classes and their traditional counterparts as classified by the NIST}
    \begin{tabular}{|l | c|}
        \hline
        Class & security comparable to \\
        \hline
        1 & AES-128 \\
        2 & SHA256 \\
        3 & AES-192 \\
        4 & SHA384 \\
        5 & AES-256 \\
        \hline
    \end{tabular} 
\end{table}

\subsection{Encryption}
\comment{ 
} 

QR encryption schemes can be based upon a multitude of different mathematical problems thought to be hard even for quantum computers.

Sadly, being thought of as secure mostly is not based upon actual rigorous proof but assumptions.
Therefor one problem that was used as a asymmetric encryption basis, the knapsack problem, was broken soon after its introduction by so-called approximate lattice reduction attacks \cite{QR_algs}.

Later iterations which include ``conjugacy search problem and related problems in braid groups, and the problem of solving
multivariate systems of polynomials in finite fields''\cite{QR_algs} have been under active research with the latter being broken after standardization and implementation \cite{QR_algs}.

Nevertheless there is an implementation of a multivariate-based scheme, called Rainbow, that is also currently a contender for standardization. But as an encryption scheme its not very suitable since the process of decrypting in multivariate based schemes requires some guessing work \cite{QR_comparison} which is not desirable in IoT environments.
An additional problem that would make rainbow unsuitable for IoT use-cases is its big 22kB public key. While private keys can rather easily be shrunk in key-generation through help of a pseudo-random-generator, thats generally not the case for large public keys.

On the other hand we have a problem that is not yet very well researched and also not much in use, but has one implementation called SIKE. This problem is based upon supersingular elliptic Curves, which are itself a modification of elliptic curve problems that should make it quantum resistant. But since this topic is not well-studied yet we are mostly left with schemes based upon the following two thought to be quantum-hard problems.

The first one is so called code-based cryptography. Here the decoder has to correct errors of data that has been seemingly randomly shuffled, but only those with access to the private key can easily `unshuffle' the data to then use special error correction codes. 
The most researched one is called McEliece and even has quite fast ($100 \mu s$) and secure implementations. 
The main problem is, that the `shuffling' is realized through $k*n$ matrices that are generally big (millions of bits) and therefor unfeasible for constrained IoT devices.

The second one will be discussed in greater detail in section \ref{QR signatures}, since it is also used as one of the main problems for signature schemes.
Those schemes are called lattice based and also have some implementation with the most famous for encryption being NTRUEncrypt\footnote{NTRU is short for N-th Degree Truncated Polynomial Ring. Also NTRUEncrypt might pop up the most, but the two main contenders for actual future use are FALCON (which also uses NTRU) and Dilithium, which uses other ring lattices}.

\subsection{Signatures} \label{QR signatures}
\comment{ 
 code-based:   
- McEliece decrypt padded message digest - try thousands of paddings - signing takes 30secs, 4mb priv/pub KEY-size -not feasbale

} 
The other pillar of cryptography, signature schemes, is what we will focus on in greater detail.
As well as in encryption schemes we can differentiate between different underlying mathematical problems. Those are pretty much the same as in encryption schemes: 
Hash based,
Lattice based,
Multivariate polynomial based,
Code based,
Super-singular isogeny based.\cite{QR_sigs}

Rainbow is the only implementation of a QR signature that is a current contender for standardization that is neither lattice nor hash based.
And as already mentioned in the previous section it is multivariate based.

Since this sparsity of alternatives we we also focus on hash and lattice based signatures in this paper.

\subsubsection{Hash Based Signatures (HBS)}\label{HBS}
Hash based signatures have their security based upon the hardness of reversing hashes or one-way functions.
The most easy one is the Lamport one time signature (OTS).\cite{QR_algs} 
That signature has essentially two private keys for every bit in the message digest. 
Let $n \in \mathbb{N}$ be the bit-length of the digest, then the secret key would be: \[k_\text{priv}= (S_{0,0},S_{0,1}) || (S_{1,0},S_{1,1})|| \dots || (S_{n,0},S_{n,1})\]
The advantage of those schemes is, that the private keys do not have to have any special characteristic that could be taken advantage of by a quantum computer to break anything.
They do have to be high entropy though, to not be easily forgeable with even a classic computer. 
These secrets are then hashed (with a one-way function $h$) and published as the public key \begin{align*}
    &k_\text{pub}= \\ &(h(S_{0,0}),h(S_{0,1})) || (h(S_{1,0}),h(S_{1,1}))|| \dots || (h(S_{n,0}),h(S_{n,1}))
\end{align*}

When a message is signed the signer just publishes the secret corresponding to every bit of the digest ($S{k,b}$ with $b$ being the bit-value in the $k$-th position ob the digest) s.t. everyone can hash that secret and see that this private keys are indeed the ones corresponding to the public key and the correct bit-value of the digest.
Signing as well as verifying are therefor rather easy operations with one disadvantage: the keys and the signature are super big.
But there are some rather easy improvements for this problem e.g. one could only sign the zeros, therefor reducing key sizes by a factor of $2$ as well as average signature sizes. To mitigate an attack that can flip digest-zeros to ones a checksum is added (that can only be decreased by flipping a one to zero, which is impossible if you do not know the pre-image (private key) of that location).
Another improvement often wrongly
\footnote{the Merkle OTS has two parts, one that is similar to the Lamport scheme (which was then improved by Winternitz) and one that uses Merkle Hash Trees, which most of the literature refers to as the Merkle Signature, but is not a predecessor of the winternitz scheme which does not use Merkle Trees \cite{Merkle_ots}} 
cited as the successor to the Merkle OTS is the Winternitz scheme (WOTS), which builds upon the same idea but uses a different (greater) basis $b$, which inturn makes the signing and verifying more computational expensive by needing to apply hashes $b$ times. The great advantage though is that the keys and signatures also decrease by a factor of $\nicefrac{b}{2}$.
This can be a great advantage for IoT applications, since time is not as valuable as storage. Therefor WOTS is actually used in practice, for example as a signature on the IOTA distributed ledger. \cite{iota_wots}

A directly visible disadvantage of those schemes (as the name implies) is that they can trivially only be used one time, since most of the private key gets public with the signature. 
A trivial countermeasure would be to append the next public keys to the message and sign them as well, but thats not a good idea in most use cases, since you might as well just use symmetric cryptography which is also considered as quantum resistant as hashes.
Another idea would be to just publish a whole lot of private keys that can than be used one by one. But thats not a super brilliant idea since signer as well as verifier need to store all these keys which is specially infeasible in IoT scenarios (that have very constrained storage).
Schemes that can be used multiple times are smartly called multiple time signatures (MTS).

A smarter approach than simply publishing $n$ public keys and storing $n$ private keys was proposed by Merkle \cite{Merkle_ots}.
His approach uses so called Merkle hash trees to make it possible to have a very small public key that can still verify $n$ signatures on the tradeoff that every signature now increases by a factor of $\log(n)$.
The idea is as shown in algorithms \{\ref{Merkle gen v1}, \ref{Merkle sig v1}, \ref{Merkle ver v1}\}.

\begin{algorithm}
\caption{GEN}\label{Merkle gen v1}
    \begin{enumerate}
        \item generate $n=2^m$ random values, those are the private keys.
        \item for every private key $k_\text{priv}^i$ generate a one-time public key $k_\text{pub}^i=h(k_\text{priv}^i)$ (until here it is similar to a trivial MTS)
        \item hash every two `neighboring' keys $k_\text{pub}^i$, $k_\text{pub}^j$ together in pairs to generate $\nicefrac{n}{2}$ new hashes $h_{ij}=h(k_\text{pub}^i, k_\text{pub}^j)$ 
        \item hash those in pairs for the next iteration and repeat until the hash-tree is complete and we only have one root hash denoted as $k_\text{pub}$
        \item publish $k_\text{pub}$ that can now be used to verify $n$ signatures
    \end{enumerate}
\end{algorithm}

\begin{algorithm}
    \caption{SIGN}\label{Merkle sig v1}
        \begin{enumerate}
            \item input message digest $M_i$
            \item sign as described for Lamport or Winternitz schemes (or other OTS schemes that generate the public key by hashing the private key): $S_i=\text{Sign}(M_i)$
            \item publish $S_i$ together with all hashes $h$ needed to iteratively generate the root hash $k_\text{pub}$. These are $m$ hashes.
        \end{enumerate}
    \end{algorithm}

\begin{algorithm}
    \caption{VER}\label{Merkle ver v1}
        \begin{enumerate}
            \item input signature $(S_i,[h_{j},h_{i+2,j+2},\dots,h_{i-k}])$ and digest $M_i$ and already known multiple use public key $k_\text{pub}=h_{0-(n-1)}$
            \item hash $S_i$ to generate $k_\text{pub}^i$
            \item hash $k_\text{pub}^i=h_i$ together with $h_j$ to generate $h_{ij}$
            \item hash the value from previous step together with the next hash given by the signature
            \item repeat step 4) until the root hash $k_\text{pub}$ should be found (thats $m$ steps in total) return True if they are equal and False otherwise
        \end{enumerate}
    \end{algorithm}

This is already very useful for IoT actors that only need to verify, less so for sensors that still need to store all $n$ private keys. 
The computational cost is higher, caused by calculating all those hashes but thats commonly worth the tradeoff. 

On the other side (the signer) we still need to store all $n$ private keys and calculate $m$ hashes every time we want to sign anything.
The second step can be skipped by also storing the hashes instead of calculating them, which additionally increases the storage needed by a factor of $2$.
But thats infeasible for most storage constrained devices.
Therefor an additional tweak was applied to this algorithm:
Instead of randomly generating each private key and storing it, we use a Pseudo-Random-Generator (PRG) together with a seed and a counter to be able to generate every private key on the fly.
We can then iteratively generate our Merkle tree and drop every node we already used to calculate the next parent hash without exceeding our RAM to generate the root hash to then publish it as the public multi time key.
For signing we can then create our private key again with the help of the PRG and again calculate all needed hashes iteratively the same way. 
But it is rather computationally expensive to recalculate the whole tree on every signature. 
That is why we should cache as many in-between hashes as possible since every already stored hash reduces the computational expenses by a factor of 2.
The verification stays the same.

This scheme is know as the eXtended Merkle Signature Scheme (XMSS) which also has some further variants and developments. \cite{QR_IoT }

Another disadvantage of schemes as described is the so-called statefulness, which means that the signer cannot just sign any message with a key after being reset, since some kind of state is needed that would be lost in a reset. \cite{QR_sigs} 
Besides that and even more impactful the verifier in an MSS needs to manage which keys/ part of the tree have already bin used, since reusing keys is imperative to the schemes security.

\subsubsection{Lattice Based Signatures (LBS)}\label{LBS}
In a stateless scheme on the other hand, all you need to sign a message is a static private key.
That brings us to the other kind of signature schemes, the ones that are more similar to traditional asymmetric crypto in the sense that they rely on a not so trivial mathematical problem that is not easily algorithmically solvable. But instead of prime factorization or elliptic curve calculation, this one seems to be hard to solve, even by a quantum computer. The problem used for most of these schemes are based upon lattices.

A lattice in this case is a high-dimensional grid with only integer values. Or to be more precise: ``An n-dimensional lattice is the set of vectors that can be expressed as the sum of integer multiples of a specific set of n vectors, collectively called the basis of the lattice—note that there are an infinite number of different bases that will all generate the same lattice'' \cite{QR_algs}
To put it mathematically we can denote a lattice $L$ as $L=\{\sum a_i*b_i : a_i \in \mathbb{Z}\}$ with $b_0,\dots,b_n$ being arbitrary basis vectors.
The mathematical problems that these schemes are based upon are the shortest vector problem (SVP) where a very short vector between to points need to be found or the Closest Vector Problem (CVP), where a lattice vector needs to be found that is closest to a given arbitrary point.
The directly arising problem though is, that to get reasonable security the basis (which serves as a private key) of the lattice needs to be in the range of megabits, which again is not ideal for our use-cases. That is why researchers developed the NTRU crypto system, that introduces certain symmetries to the lattice structure s.t. the key sizes can be much smaller while lowering the security only slightly.\cite{QR_algs,QR_comparison}
These new schemes are not only resistant to quantum attacks but also improve efficiency compared to traditional cryptography by having speed improvement by a factor of 10-100.\cite{QR_sigs}
Sadly these lattice structures where vulnerable through lattice reduction techniques to Chosen Ciphertext Attacks (CCA) in the case of encryption schemes. But that was fixed with the introduction of a special padding scheme that made these attacks impossible but also increased the key-lengths. \cite{QR_algs}
In the case of signature schemes (like NTRUSign) the problem is even more severe.
The signature works by first mapping the message to a vector and then signing by solving the CVP for this vector. The problem is, that this procedure leaks information about the private key s.t. it was shown to be practically broken after only around 400 signatures.  To mitigate that issue the signer does not give the actual closest lattice vector, but a lattice vector that is close enough by a certain measure, but not necessary the closest. Therefor the leaked information is nearly neglectable and the signature and private key secure for around a billion signatures, although it is still advised to change the private key after around 10 million signatures. 
That is totally feasible compared to some MTS mentioned before since in many cases 10 million signatures is a whole lot.\cite{QR_IoT}

Actual lattice based implementation that were proposed in 2017 are GPV, GLP and BLISS. But now there are newer and better implementations like FALCON that will be discussed in section \ref{falcon}.

\subsubsection{Comparison of HBS and LBS and Statefulness versus statelessness}
At the time \cite{QR_sigs} was written, one the most common HBS was SPHINCS, which has not changed much, although it got a major update and has quite a few variants.
The most prominent LBS scheme was called BLISS.
Suhail et al. then compared these two implementation and measured their performance.
BLISS (the LBS) was evaluated on a common IoT processor with ARMs M4 architecture while SPHINCS (the HBS) was evaluated on a intel XEON server processor. Still BLISS performed considerably better with exception of the key sizes. Results of their measurements are shown in table \ref{t:sphincsVSbliss}.

\begin{table}[]
    \centering
    \caption{Measurement results of comparing BLISS on M4 with SPHINCS on intel XEON as of \cite{QR_sigs}}
    \label{t:sphincsVSbliss}
    \begin{tabular}{|r | c  c|}
        \hline
        Metric & SPHINCS results & BLISS results\\
        \hline
        Signature clock-cycles & 50 million & 5.9 million \\
        Verification clock-cycles & 1.6 million & 1 million \\
        Public key size & 1KB & 7KB \\
        Private key size & 1KB & 2KB \\
        Signature size & 41KB & 960 bytes \\
        \hline
    \end{tabular}
\end{table}

We can therefor compare and somewhat conclude the pros and cons of LBS compared to HBS.
First of all it is to say, that HBS is already very well studied and bases its security upon already well established and praxis-tested Problems (hashing), while lattice based Security is still quite new and in active research. It bases its security upon the CVP (or SVP) which itself is not studied that well, and a few vulnerabilities have already been shown \cite{QR_sigs}.
Another comparison, less about the underlying mathematical problems, but more from an applicational standpoint, is also from interest: statefulness.
We already know what stateful means, but Suhail et al. summarizes it rather well: 
``stateful digital signature scheme necessitates the maintenance of the updated nonrepeated secret key upon each signature generation process. It is essential to keep track of nonrepeated key pairs, failing which will result in the degradation of the security of the cryptographic scheme'' 
We can already see that this would be a problem in many use-cases.
There are HBS schemes, like the above SPHINCS, that found a way to to make themselves stateless, but that comes with its own downsides, like in this case greatly increasing the key-generation expenses (compare table \ref{t:sphincsVSbliss}) , which is caused by using keys in a random order therefor making BDS\footnote{compare \cite{BDS_optim}} optimization no longer applicable. 

It can therefor be concluded that stateful schemes are great in processor power/time constrained use cases while stateless schemes trade computing power for storage used and are therefor better in memory constrained use-cases. \cite{QR_sigs}

\section{QR Signatures in IoT}

We now have an overview of what kind of QR signature schemes exist with some focus on the two most promising underlying mathematical structures, hashes and lattices. In the following sections we will even more focus on what kind of actual implementation exist and which of them are feasible in an constrained environment.

\subsection{Performance Metrics in IoT}
\comment{ 

'' small sized public key, small digital signature and a range of supported hash output sizes is recommended''\cite{QR_Iot_Lattice}

- „signature and/or key sizes to running times and memory consumption to energy consumption „
- „From the software benchmark perspec- tive, the runtime of key generation, signing, and verification processes whereas from the hardware perspective, CPU cycles, key size, signature size, and energy consumption are among the targeted evaluation metrics. In general, the parameter sets are highly dependent on the underlying construction of a particular scheme.“ \cite{QR_sigs}

} 

First of all we need to establish what metrics we need to consider to evaluate which implementation are better suited and which are ill suited.
As mentioned in section \label{bg:iot} we have constraints in different fields. Those are primarily available energy therefor computing time and power as well as storage.
The storage itself can also be split into read only memory (ROM) and writeable memory (RAM).
Most operating systems (e.g. RIOT) split the available RAM again to expose two kinds of memory structures: stack and heap. These are often relevant since the stack and heap are used for different purposes, but since those are resizable at build time we wont focus on the differentiation between stack and heap.
Another memory differentiation that might be more relevant for future differentiation between stateful and stateless schemes is between persistent and volatile memory. 
For stateful schemes we need persistent writeable memory and probably quite a lot of it while stateless schemes typically require more computing time for every signature. Which brings us to the main metrics:

For each algorithm (GEN, SIGN,VER) we are interested in RAM/cache usage, execution time/energy consumption.
The significance of SIGN and VER performance is pretty obvious, but the GEN is also not to be forgotten since we have shown that most schemes either need to switch the keys because they leak over time or are simply just MTS i.e. have to switch keys after a fixed amount of signatures, this amount is also influenced by available storage/time as we have shown in section \ref{HBS} with the Merkle tree based signatures.
We are also interested in different sizes that need to be either stored or transmitted via the network, these are signature as well as private and public key sizes. In many schemes these sizes have been unusually big \cite{QR_comparison}.
The overall needed ROM for any algorithm is also from great interest since this is needed to say whether a scheme is applicable for different classes of IoT nodes as mentioned in section \ref{bg:iot}.

Another thing regarding these classes that we might want to mention is, that there is no quantum computer with even nearly sufficiently sized quantum registers yet and probably will not be in only a few years. Therefor it is probable that when these quantum computers exist, the IoT and their hardware will also have gotten much better. Still, the right signature schemes are needed, but it might be okay to be a few kilobytes larger than these classes.  

\subsection{comparison of different signatures}
\comment{ 
Hash-Based Sphincs promising since stateless, but many parameters to set \cite{QR_IoT_Energy}
 
as of \cite{QR_comparison} only schemes (out of ~50) with < 4kbit: SIKE and Round5 
} 

Since we now know what to look out for we can now compare results of measurements from different QR signature implementations.
In this comparison we already focused on schemes that could be relevant for the IoT (i.e. skipped schemes that had way to much memory/cpu usage or are to old, e.g. in \cite{QR_comparison} the only schemes that had below 4kB keys where SIKE and Round5, both of which did not make it to the NIST finalist).
The measurements seen in tables \ref{t:stack_comp} and \ref{t:clockcycles_comp} were performed by \cite{QR_comparison},\cite{Energy_comp} and \cite{update_sign}

\begin{table}
    \caption{Comparison of stack usages for different schemes and their operations ( - means that it has not bean measured while / means not applicable)}
    \label{t:stack_comp}
    \centering\begin{tabular}{| r | c c c |}
        \hline
        Implementation name                     & GEN (bytes) & SIGN (bytes) & VER (bytes)\\
        \hline
        Dilithium-3 \cite{QR_Iot_Lattice}       & 50k       & 86k   & 54k\\
        2021 Dilithium(dyn)\cite{update_sign}   & -         & 52k   & 36k\\
        2021 Dilithium(sta)\cite{update_sign}   & / \footnote{in the case of static Dilithium the keys where precomputed and directly stored in flash} & 35k   & 19k\\ 
        qTESLA-1 \cite{QR_Iot_Lattice}          & 22k       & 29k   & 23k\\
        qTESLA-3 \cite{QR_Iot_Lattice}          & 43k       & 28k   & 45k\\
        Falcon-5  \cite{QR_Iot_Lattice}         & 120k      & 120k  & 120k\\
        2021 FALCON \cite{update_sign}          & -         & 42k   & 4.7k\\
        \hline
    \end{tabular}
\end{table}

\begin{table}
    \caption{Comparison of clock cycles needed for the operations of different implementations, performed on ARM M4 chip which was clocked at 168Mhz therefor 10 million clock cycles equal roughly 60ms. Each value is a million clock cycles}
    \label{t:clockcycles_comp}
    \centering\begin{tabular}{| r | c c c |}
        \hline
        Implementation name                     & GEN           & SIGN         & VER \\
        \hline
        Dilithium-3 \cite{QR_Iot_Lattice}       & 2.3           & 8.3          & 2.3 \\
        Dilithium-3 \cite{Energy_comp}          & 2.1           & 7.2          & 2.1 \\ 
        2021 Dilithium(dyn)\cite{update_sign}   & -             & 29           & 3.4\\
        2021 Dilithium(sta)\cite{update_sign}   & -             & 8            & 1.5\\
        qTESLA-3 \cite{QR_Iot_Lattice}          & 30            & 11           & 2.2\\
        Falcon-5 \cite{QR_Iot_Lattice}          & 365           & 165          & 1\\
        2021 Falcon  \cite{update_sign}         & -             & 75           & 1 \footnote{after optimizations these could be improved by further 43\% \cite{falcon_micro_impl}}\\
        \hline
    \end{tabular}
    
\end{table}

Unfortunately i could not find any reliable data about compiled code size \footnote{And i did not have enough resources left to measure it on my own, this could be part of future comparison research} (i.e. ROM usage) except for Dilithium and FALCON which is shown in table \ref{t:flashsize_comp} which is an important parameter, but the official NIST competition reference implementations \cite{nist_finalists_website} can give us a rough idea if we have a look how big the source code files are. This is shown in table \ref{t:codesize_comp}.

\begin{table}
    \caption{uncompiled code size of reference implementations (different security levels (like Dilithium-3 and Dilithium-5) do not have any static changes reflected in code size)}
    \label{t:codesize_comp}
    \centering\begin{tabular}{| r | c |}
        \hline
        Scheme & Size\\
        \hline
        FALCON & 372KB \\
        Dilithium & 270KB\\
        SPHINCS+ & 180KB\\

        \hline
    \end{tabular}
\end{table}

\begin{table}
    \caption{Flash sizes)}
    \label{t:flashsize_comp}
    \centering\begin{tabular}{| r | c |}
        \hline
        Scheme & Size \\
        \hline
        FALCON & 57KB \\
        2021 Dilithium (Dyn) & 11KB\\
        2021 Dilithium (Sta) & 26KB\\
        \hline
    \end{tabular}
\end{table}

\begin{table}
    \caption{Comparison of key and signature sizes}
    \label{t:key_sig_comp}
    \centering\begin{tabular}{ | r | c c | }
        \hline
        Scheme & public key & signature \\
        \hline
        SPHINCS     & 1KB   & 43KB \\
        Dilithium-3 & 1.4KB & 2.7KB\\
        FALCON-1    & 900B  & 690B\\
        FALCON-5    & 1.7KB & 1.3KB\\
        \hline
        ECDSA       & 64B   & 64B\\
        \hline
    \end{tabular}
\end{table}

After comparing all these signatures and keeping in mind that only Rainbow (implemented in python, not quite ready for IoT), Falcon and Dilithium became successful finalists \cite{nist_finalists_website}, we can conclude that there are different schemes with their own strengths and weaknesses.
But in conclusion we can see that Dilithium shines on the signing side while FALCON performs by far the best when it comes to verification, which is especially useful on actor nodes or other nodes that primarily need to verify received data e.g. for updates.
While Dilithium would be better in use-cases where sensitive data has to be signed.
But, all these schemes are stateless and both Dilithium and FALCON (as well as qTesla) are lattice based schemes, the last finalist, Rainbow is code based, and as we can also see from the enormous signature size SPHINCS is the only hash based contender, but did not make it to the finalists.
It was also observed that FALCON and Dilithium have the best security against quantum computers and annealers to computational expenses ratio \cite{springer_security_analyses}. We will therefor focus on these schemes in the following sections.

\comment{ 
} 

\comment{ 
wahrscheinlich skippen da mir da hintergrund zu braid groups etc fehlt
} 

\comment{ 
not in the endgame but also not broken afaik
} 

\section{FALCON and Dilithium}\label{falcon}
\comment{ 
falcon-512 (L1):
pubk/sig 897/690 bytes (dil3: 1472/2701 ecdsa: 64)
keygen: 182m clk , 118mJ (dil3: 2.3m / 1.7mJ ecdsa 5mJ)
sign/ver: 23.5/0.345 mJ (dil3 5mJ/1.7mJ ecdsa 4mJ)

falcon-1024 (L5):
pubk/sig 1793/1330 bytes
keygen: 380m clk , 232mJ
sign/ver: 45.5/0.69 mJ
\cite{Energy_comp}

} 
Since we are now focussing on lattice based algorithms it would make sense to describe these again in further detail.
We already know (compare section \ref{LBS}) that a lattice is essentially a high dimensional grid (over a residue field). 
We also know that the underlying problem used for the schemes security is some form of CVP (or SVP), but we will now show how an actual cryptographic algorithm is based upon these problems.

The main idea is that these lattices have a nearly infinite amount of bases, some of which make it easy to solve these problems, others make it hard.
We can now construct a lattice with two equal bases, one with rather short vectors, that make it easy to solve these problems and one with arbitrary long vectors, that make it hard.
The reason behind a basis being better than another is complicated but in can be greatly illustrated with an algorithm that is used to solve this CVP which is called Babai's rounding technique.\cite{CVP_algorithms}

\begin{algorithm}
    \caption{Babei's rounding technique}\label{alg:babei}
        \begin{enumerate}
            \item input lattice basis $B$ and vector $v$ for which we search the closest (in practice a close enough) lattice vector
            \item transform basis of $v$ into $B$ (since $v$ uses a standard basis this is a trivial step, the problem is, that the probability that our new representation uses integer values is neglectable)
            \item round our new representation to the next integer values. Return these.
        \end{enumerate}
    \end{algorithm}

This algorithm can be seen in \ref{alg:babei} and we can now see that using short vectors our rounding brings us to an actually close vector, while a long basis wont.
Just imagine the most trivial case, a one dimensional lattice that consist of all values in $\mathbb{Z}_99$. We now want to solve the CVP for the vector $v= \{24.68\}$.
If we now have a short basis (e.g. \{2\}) this algorithm for finding a close vector will represent $v$ as $12.34*\{2\}$ and then round to $12$ which would give us $12*\{2\}=\{24\}$ which is indeed a close vector, although not the closest, which would have been $\{25\}$ as we know.
On the other hand, if we have a long basis (e.g. \{14\}) this algorithm for finding a close vector will represent $v$ as $1.72\dots*\{7\}$ and then round to $2$ which would give us $2*\{14\}=\{28\}$ which is not a close vector.
Of course this example would not make any sense in a practical environment but it highlights why a short basis is good for solving CVP while a long one is not.
But both bases are equally useful in verifying that a vector is indeed a lattice vector by simply checking if it can be represented by an integer multiple of our basis.
With these properties we can now create a first simple lattice based signature scheme which is given by \{\ref{alg:lattice gen v1}, \ref{alg:lattice sign v1}, \ref{alg:lattice ver v1}\}.

\begin{algorithm}
    \caption{Lattice GEN}\label{alg:lattice gen v1}
        \begin{enumerate}
            \item Choose an arbitrary $n$-basis consisting of short vectors. This is the private key.
            \item Choose any other basis consisting of long vectors for the same lattice. This is the public key. 
        \end{enumerate}
    \end{algorithm}

\begin{algorithm}
    \caption{Lattice SIGN}\label{alg:lattice sign v1}
        \begin{enumerate}
            \item input message digest $M_i$ and already known private key being a lattice basis $b_\text{priv}$
            \item genrate random seed $r$
            \item compute $u=h(M_i,r)$
            \item map $u$ to a $n$-dimensional vector $v_u$
            \item solve CVP for $v_u$ by taking advantage of our small basis $b_\text{priv}$ (similar to \ref{alg:babei}, but more complicated in practice) and assign the found vector to $p$
            \item $s=p-u$; return signature $(r,s)$
        \end{enumerate}
    \end{algorithm}

\begin{algorithm}
    \caption{Lattice VER}\label{alg:lattice ver v1}
        \begin{enumerate}
            \item input signature $(r,s)$ and digest $M_i$ and already known public key being another lattice basis $b_\text{pub}$
            \item abort if $s$ is not small (the vector is not close to the given digest vector)
            \item compute $u=h(M_i,r)$
            \item map $u$ to a $n$-dimensional vector $v_u$
            \item check whether $v_u$ is an actual lattice vector given by your basis $b_\text{pub}$ and return result
        \end{enumerate}
    \end{algorithm}

\subsection{FALCON}

This is however not implemented one-by-one into FALCON, because of two main issues.

The first one is that we need a high dimensional lattice (to not be prone to LLL Lattice Reduction) and an arbitrary basis lies in $\mathcal{O}(n^2)$ which would result in unfeasible Megabyte-sized keys.
The solution to this problem is to not use any arbitrary bases but introduce some kind of symmetry that makes the keys a lot smaller. This symmetry is introduced with the already mentioned NTRU-Lattice, which upscales only a few basis vectors to a high dimensional basis by applying specific rotations.

The second problem (compare section \ref{LBS}) is the continuous leakage of our private key from signature to signature that arises through usage of algorithm \label{alg:babel}. That is because in this algorithm the shape of the closest vector is directly related to the shape of the private basis.
To mitigate this issue a random sampling is introduced. It is called the GPV sampler \cite{falcon_easyier_read} and it works by not rounding to the closest integer but randomly rounds up or down.
FALCON also improved this sampling process by applying Fast Fourier Sampling which sped up the process by orders of magnitude but as later figured out introduced a new vulnerability. 

The new problem lies in the use of the floating point unit for the fast fourier calculations. This is not only a problem since many IoT devices do not have a floating point unit, but it also introduced a side-channel attack vulnerability.
Floating Point Arithmetics (FPA) are known to be prone to side-channel attacks however in this case it has a somewhat severe impact.

FALCON was side channel attacked by measuring electromagnetic radiation created by the FFT inside the sampler which uses the floating point unit. Such that the attacker could reconstruct the private key after only 10k measurements \cite{falcon_side_channel}. Therefor another fork af falcon, zalcon was introduced which ditches the FPA (also good for other iot devices) and uses NTT instead of FTT \cite{zalcon}. However this NTT seems to be even more vulnerable to this attack. \cite{falcon_side_channel} 

A fix for this vulnerability though could be to introduce a masking s.t. power consumption etc. is more randomized. This would be a rather standard approach to mitigate this FPA Side-Channel Attack.

Besides this timing attack another team \cite{bearz} created a fault attack that could also retrieve the private key, but they propose a countermeasure for this is detecting the fault attack by verifying the own signature, which can take further advantage of FALCONs fast (~3ms) verification time.

To conclude FALCON would be a good contender for class C2 Devices and if only the verification part is needed it could also be used in C1 devices, which no other QR signature scheme can do. 
It is not super feasible in real-time scenarios for sensor networks though, since signing data can take about one whole second (compare table \ref{t:clockcycles_comp}).

If we also take energy consumption into consideration (see table \ref{t:falconVSdil}) we can see that FALCON really shines in verifying messages and is even super efficient on the highest security level. But the key generation takes a lot of energy which would be bad in battery operated environments. But even with a small 600mAH Battery that would still be around 34 thousand key generations before the battery is drained and since a single key can safely be reused for 10 million signatures that would result in 340 billion signatures made possible by pretty much the smallest available LiPO-cell. But a device would not even get close to this many signatures since signing with FALCON is way more energy intense then its competitor Dilithium.

\begin{table*}[]
    \caption{Direct comparison of FALCON and Dilithium on M4. \cite{Energy_comp}}
    \label{t:falconVSdil}
    \centering\begin{tabular}{ | r | c c c c c |}
        \hline
        Scheme & public key size & signature size & GEN Energy & SIGN Energy & VER Energy \\
        \hline
        Dilithium-3 (L2)    & 1472B & 2701B & 2.3mJ & 5mJ   & 1.7mJ \\
        FALCON-512 (L1)     & 897B  & 690B  & 118mJ & 23mJ  & 0.3mJ \\
        FALCON-1025 (L5)    & 1793B & 1330B & 232mJ & 35mJ  & 0.7mJ \\
        \hline
        ECDSA               & 64B   & 64B   & 1.7mJ & 4mJ   & 4mJ \\
        \hline
    \end{tabular}
\end{table*}

\subsection{Dilithium}
\comment{ 
FALCON reaches small pubk and sig since usage of gaussian sampling, - also makes signing hard, dilithium doesnt

runs in constant time - no timing attacks possible \cite{Dilithium}

} 

Dilithium's main idea is similar to the one used by FALCON, with the main difference being that it ditches the gauss sampler.
This sampler is the main reason FALCON can achieve such small public keys and signatures, but it also makes the implementation a lot complexer and the signature step much harder, as we have seen in the previous section. 
Since Dilithium ditches this step for another more trivial step, it makes signing much faster and easier while also reducing the code sizes needed to store this scheme on any device, making it possible to be only a small part even on a C0 device. However neither signing nor verifying would work on a C1 device, while all would work on a C2 Device which is already good.
Another advantage of ditching this sampling is that it enabled the Dilithium team to focus on making the whole scheme constant-time. This inturn makes timing attacks as these that FALCON is vulnerable to impossible by design. \cite{Dilithium}


\section{Conclusion}
\comment{ 
- of course no protection against side channel etc 
- quantum fast evolving, active field of research
- smart home, smart campus, smart city
- quantum key distribution
\cite{QR_comparison}

- similar performance as traditional and also implemented in TSL variant \cite{falcon_and_dilithium}

Dilithium when signing and verification often needed
Falcon if only very since hard part is gaussian smapling which can be omitted on ver
} 

To conclude we can see that there are two NIST finalist with promising performance even for the IoT, some performance improvement for specific platforms are still possible (like Dilithiums avx-2 vector optimizations), but even without those we have performance competitive with more traditional schemes like ECDSA.
Both of them are lattice based, since the most used alternatives are hash based, which results in very big signatures by design.

We had a look at multiple different schemes and their underlying mathematics and found that HBS are easier to understand and proof their security with only assuming hardness of pre image attacks for hashes. But we also found that their signatures get unfeasible big for IoT scenarios, therefor needing lattice based signatures, that offer better signature length performance on the downside of building their security on a less researched mathematical problem.

Still the two LBS schemes that were further researched, FALCON and Dilithium would both fit on C2 IoT nodes and perform reasonably well on them.
Besides memory constraints and FALCONs performance demands during signing due to the hardness of gauss sampling, they both have performance comparable or even better to more traditional non-QR schemes \cite{falcon_and_dilithium}. If we only want to verify messages FALCON would be an even better performer, since most of its hard part can be omitted, and signature sizes, public keys and verification hardness are very small.

But FALCON still faces some issues on the signing side, since it is quite demanding and not constant-time which inturn poses vulnerability against side channel attacks.
Other attacks like fault injection etc still have to be evaluated for concrete implementations on concrete devices.
Also the quantum-resistance is based on the CVP, and quantum computing in general is still an active field of research and rapidly evolving.

But with sufficiently large quantum computers on the horizon we need to make sure to quantum proof our smart homes, campuses and cities before it is to late. And FALCON as well as Dilithium seem to be reasonable contenders for that.

\bibliographystyle{IEEEtran}
\bibliography{lit.bib, lit_specific.bib}

\end{document}